

\magnification=\magstep1

\def \w {\omega}
\def \H {{\cal H}}
\def \t {\theta}
\def \P {\Psi}

\pageno=0
{\nopagenumbers
\line{\hfil September, 1993}
\vskip 3 cm
\centerline {\bf Quantum Dilogarithm}
\vskip 2 cm
\centerline {L.D. Faddeev$^{1,2,}$\footnote{$^\dagger$}
{Supported by the Russian Academy of Sciences and Academy of Finland},
R.M. Kashaev$^{1,}$\footnote{$^*$}
{On leave of absence from St. Petersburg Nuclear Physics
Institute, Gatchina, St. Petersburg 188350, Russia}$^{,\dagger}$}
\vskip 1 cm
\centerline {\it $^1$Research Institute for Theoretical Physics,}
\centerline {\it P.O.Box 9 (Siltavuorenpenger 20C), SF-00014 University
of Helsinki, Finland}
\vskip 0.5 cm
\centerline{\it $^2$St. Petersburg Branch of the Steklov Mathematical
Institute,}
\centerline{\it Fontanka 27, St. Petersburg 191011, Russia}
\vskip 4 cm
\line{\bf Abstract \hfil}

A quantum generalization of Rogers' five term, or ``pentagon''
dilogarithm identity is
suggested. It is shown that the classical limit gives usual Rogers'
identity. The case where the quantum identity is realized in finite
dimensional space is also considered and the quantum dilogarithm
is constructed as a function on Fermat curve,
 while the identity itself is equivalent to the
restricted star-triangle relation introduced by Bazhanov and Baxter.
\vfil
\eject}

\beginsection{1. Introduction}

Recently, the dilogarithm function appeared in the
study of the two dimensional  quantum conformal field theory (CFT)[1] and
solvable lattice models both in two [2] and three dimensions [3].
Particularly, the effective central charges of some rational CFT can
be expressed as finite sums of Rogers' dilogarithm function with
the arguments being solutions for certain algebraic relations
(e. g. thermodynamical Bethe anzats relations).

 People use the dilogarithm function in two forms, $L_2(x)$ and $L(x)$,
introduced by Euler and Rogers, respectively. Euler's dilogarithm
for any complex $x$ can be defined by the following integral:
$$
L_2(x)=-\int_0^x{\log(1-z)\over z}dz,                     \eqno(1.1)
$$
while Rogers' differs by the addition of an elementary function
$$
L(x)=L_2(x)+\log(1-x)\log(x)/2.                            \eqno(1.2)
$$
Among the various properties of these functions the most
profound one is (probably) the five term or ``pentagon'' identity for
$L(x)$, proved by Rogers [4]:
$$
L(x)+L(y)-L(xy)=L({x-xy\over1-xy})+L({y-xy\over1-xy}),      \eqno(1.3a)
$$
which in terms of Euler's dilogarithm looks like
$$
L_2(x)+L_2(y)-L_2(xy)=L_2({x-xy\over1-xy})+L_2({y-xy\over1-xy})
+\log({1-x\over1-xy})\log({1-y\over1-xy}).      \eqno(1.3b)
$$

 On the other hand, in the context of integrable lattice quantum field
theory models and quantum groups the
problem of constructing of the Virasoro algebra on the lattice has been
suggested and partially solved [5], [6]. The
following transcendental function plays the significant role in the latter:
$$
\P(x)=\prod_{n=1}^\infty(1-x\t^n),\quad |\t|<1,                \eqno(1.4)
$$
where $x$ is any complex variable and $\t$, a fixed complex
parameter, lies within a unit disk\footnote{$^\star$}{In [6] this function
has been defined using a slightly different argument and denoted as
$S(x)$, with $\t$ being $q^2$ of [6]. In fact, the precise relation between
the two definitions is $\P(x)=S(-\t^{1/2}x)$.}.
Considering $\t$ as a quantization parameter, the function $\P(x)$
can be
interpreted as a ``quantum'' Euler's dilogarithm in the following sense.
Let $\t=\exp(\epsilon)$ for ${\rm Re}(\epsilon)<0$ and perform the limit
$\epsilon\to0$. Then, it is easy to see that
$$
\P(x)={1\over\sqrt{1-x}}\exp(L_2(x)/\epsilon)(1+ {\cal O}(\epsilon)).
                                                            \eqno(1.5)
$$
Motivated by relation (1.5) we address the following
problem: What is the proper generalization of Rogers'
 ``pentagon'' identity (1.3) in terms of $\P(x)$? The purpose of this paper
is to answer this question.

In Section 2 a quantized ``pentagon'' identity is suggested and
it is shown that in the classical limit it is reduced to (1.3).
In Section 3 the case where $\t$ is a root of unity is considered.
The quantum dilogarithm for such a $\t$ is a function on Fermat curve,
while the classical limit of the ``pentagon'' identity again gives (1.3).
In the summary the main formulae are enumerated with a little discussion.

\beginsection{2. Quantized Rogers' Identity}

Let $\hat U$ and $\hat V$ be two operators, satisfying the Weil algebra:
$$
\hat U\hat V=\t \hat V\hat U.                            \eqno(2.1)
$$
In [6] two relations were proved for function
(1.4) with the operator arguments:
$$
\P(\hat U)\P(\hat V)=\P(\hat U+\hat V),               \eqno(2.2)
$$
and
$$
\P(\hat V)\P(\hat U)=\P(\hat U-\hat U\hat V+\hat V).    \eqno(2.3)
$$
By applying (2.2) twice to the right hand side of (2.3), the latter can be
rewritten as follows:
$$
\P(\hat V)\P(\hat U)=\P(\hat U)\P(-\hat U\hat V)\P(\hat V). \eqno(2.4)
$$
We shall argue that (2.4) is a quantum generalization of (1.3) we are
looking for. Namely, we shall show that (1.3) follows from (2.4) in the
appropriate classical limit $\hbar\to 0$. To do it, we shall write
(2.4) in terms of symbols of operators (see for example [7]).

One can realize operators $\hat U$, $\hat V$ through a pair of coordinate and
momentum Hermitian operators, $\hat q$ and $\hat p$, satisfying the Heisenberg
commutation relations:
$$
[\hat q,\hat p]\equiv\hat q\hat p-\hat p\hat q=i\hbar.      \eqno(2.5)
$$
Expressions for $\hat U$ and $\hat V$ have the forms:
$$
\hat U=x\exp(\alpha\hat q),\quad \hat V=y\exp(\beta\hat p), \eqno(2.6)
$$
where $x$ and $y$ are arbitrary complex variables, while complex parameters
$\alpha$ and $\beta$ are such that:
$$
\t=\exp(i\alpha\beta\hbar).                                \eqno(2.7)
$$
In Hilbert space, $\H$, where operators $\hat q$ and $\hat p$ act, there
are two basis sets, $\{|q\rangle \}$ and $\{|p\rangle \}$, which diagonalize
$\hat q$ and $\hat p$, respectively. We have the following scalar products:
$$
\langle q|q'\rangle =\delta(q-q'),\quad \langle q|p\rangle
=\exp(iqp/\hbar),\quad
\langle p|p'\rangle =2\pi\hbar\delta(p-p'),
\eqno(2.8)
$$
while the corresponding decompositions of the unity are
$$
1=\int_{-\infty}^{+\infty}|q\rangle dq\langle q|
=\int_{-\infty}^{+\infty}|p\rangle {dp\over 2\pi\hbar}\langle p|.
\eqno(2.9)
$$
The ``q-p'' symbol of any operator $\hat A$, acting in $\H$, is defined as
follows:
$$
(\hat A)(q,p)={\langle q|\hat A|p\rangle \over \langle q|p\rangle }.
         \eqno(2.10)
$$
The symbol of the product of two operators $\hat A$ and $\hat B$ is
given by
$$
(\hat A\hat B)(q,p)=\int_{-\infty}^{+\infty}\int_{-\infty}^{+\infty}
{dq'dp'\over 2\pi\hbar}\exp(i(q-q')(p'-p)/\hbar)
(\hat A)(q,p')(\hat B)(q',p).
                                                            \eqno(2.11)
$$
The symbols of the operators $\P(\hat U)$, $\P(\hat V)$, and
$\P(-\hat U\hat V)$, with
$\hat U$ and $\hat V$ given by (2.6), have the form
$$
(\P(\hat U))(q,p)=\P(x_q),\quad
(\P(\hat V))(q,p)=\P(y_p),                            \eqno(2.12a)
$$
$$
(\P(-\hat U\hat V))(q,p)=1/\P(x_qy_p),                \eqno(2.12b)
$$
where
$$
x_q=x\exp(\alpha q),\quad y_p=y\exp(\beta p).               \eqno(2.13)
$$
Formula $(2.12b)$ can be proven through the use of the following power
series expansions for $\P(x)$ and $1/\P(x)$:
$$
\P(x)=\sum_{n=0}^\infty\t^{n(n+1)/2}(-1)^n x^n/(\t)_n,\quad
1/\P(x)=\sum_{n=0}^\infty \t^nx^n/(\t)_n,\eqno(2.14)
$$
where
$$
(\t)_0=1,\quad (\t)_n=(1-\t)(1-\t^2)\ldots(1-\t^n),\quad n\ge1.  \eqno(2.15)
$$
Now, we can calculate the symbols of both sides of the quantum identity
(2.4), using (2.10) -- (2.12). The result has the form:
$$
{\P(x_q)\P(y_p)\over\P(x_qy_p)}=
\int_{-\infty}^{+\infty}\int_{-\infty}^{+\infty}{dq'dp'\over2\pi\hbar}
\exp(i(q-q')(p'-p)/\hbar)\P(x_{q'})\P(y_{p'}),                \eqno(2.16)
$$
where $x_q$ and $y_p$ are defined in (2.13).
In the classical limit $\hbar\to0$ we can use (1.5) with
$$
\epsilon=i\alpha\beta\hbar \eqno(2.17)
$$
 and apply the stationary phase method
for evaluation of the integrals in (2.16). Then we get
in the leading order
$$
\eqalign{
\exp({1\over\epsilon}[&L_2(x_q)+L_2(y_p)-L_2(x_qy_p)])\cr
=&\exp({1\over\epsilon}[L_2({x_q-x_qy_p\over1-x_qy_p})+
L_2({y_p-x_qy_p\over1-x_qy_p})+\log({1-x_q\over1-x_qy_p})
\log({1-y_p\over1-x_qy_p})]),\cr}
                                                        \eqno(2.18)
$$
which is $(1.3b)$ in the exponentiated form.

\beginsection{ 3. Quantum Dilogarithm at $\t^N=1$}

When $|\t|=1$, definition (1.4) of the quantum dilogarithm needs
qualification because the infinite product is not convergent. In this
section we restrict ourselves to a particular case, $\t^N=1$, for some
integer, $N\ge2$, and properly modify the formulae of Section 2.

Let $\t=\w$, where $\w$ is a primitive $N$-th root of unity. We choose
$\w$ together with its square root $\w^{1/2}$ as
$$
\w=\exp(2\pi i/N), \quad \w^{1/2}=\exp(\pi i/N).    \eqno(3.1)
$$
In this case, operators, satisfying algebra (2.1), can be realized as
$N$-by-$N$ matrices, since their $N$-th powers are the central elements.
Let us fix them equal to the minus identity operator :
$$
\hat U^N=-1,\quad \hat V^N=-1.                            \eqno(3.2a)
$$
As a consequence of this choice we have also
$$
(-\hat U\hat V)^N=-1.                               \eqno(3.2b)
$$
The spectrum of any operator, $\hat A$, whose $N$-th power is $-1$, is given by
$N$ distinct numbers
$$
\w^{n-1/2}, \quad n=0,1,\dots,N-1,                   \eqno(3.3)
$$
so, all operators $\hat U$, $\hat V$, and $-\hat U\hat V$
have one and the same spectrum (3.3). Index $n$ in (3.3) can be
considered$\pmod N$, i. e. as an element of $Z_N$.

Before specification of the quantum dilogarithm, following
[8], for complex variables $a,b,c$, constrained by Fermat equation
$$
a^N+b^N=c^N,            \eqno(3.4)
$$
define the function
$$
w(a,b,c|0)=1,\quad w(a,b,c|n)=\prod_{j=1}^nb/(c-a\w^j),\quad n\ge1, \eqno(3.5)
$$
which depends on $n$ only$\pmod N$ due to relation (3.4):
$$
w(a,b,c|n+N)=w(a,b,c|n).                                  \eqno(3.6)
$$
Quantum dilogarithm $\P(\hat A)$, depending on an operator argument,
$\hat A$ with spectrum (3.3), can be defined as an operator,
commuting with $\hat A$, and with spectrum given by
$$
\P(\w^{n-1/2})=\P(\w^{-1/2})w(a,b,c|n),                     \eqno(3.7)
$$
where $\P(\w^{-1/2})$ is a non-zero complex factor to be specified later.
Using definition (3.7), it is easy to derive the ``functional''
relation on $\P(\hat A)$:
$$
\P(\w^{-1}\hat A)\P(\hat A)^{-1}=(c-\w^{1/2}a\hat A)/b,   \eqno(3.8)
$$
which completely determines operator $\P(\hat A)$ up to a complex factor.
Now, we formulate our statement.

\noindent
{\bf Theorem}. {\it The following ``quantum pentagon'' identity holds}:
$$
\P(\hat V)\P'(\hat U)=\P''(\hat U)\P'''(-\hat U\hat V)\P''''(\hat V),\eqno(3.9)
$$
{\it where $\P^\#$ ($\#$ symbolizes any number of dashes from zero to
four) are defined by (3.8) with $a^\#,b^\#,c^\#$ satisfying (3.4), and}
$$
\eqalign{
yy''=y'y'''',\quad y'=y''y''',\quad x'''&=xx',\cr
x''=x'y'''',\quad x''''&=xy'',\cr}                      \eqno(3.10a)
$$
{\it for}
$$
x^\#=\w^{1/2} a^\#/b^\#,\quad y^\#=c^\#/b^\#.                \eqno(3.10b)
$$

\noindent
To prove the theorem, note that the combination
$$
\P''(\hat U)^{-1}\P(\hat V)\P'(\hat U)\P''''(\hat V)^{-1}  \eqno(3.11)
$$
commutes with operator $-\hat U\hat V$, provided relations (3.10) hold.
Therefore (3.11) is some function of $-\hat U\hat V$. It is a simple task
to show that this function satisfies (3.8). So, relation (3.9) holds up to some
complex factor. The latter can be put to unity after the proper choice of
$\P^\#(\w^{-1/2})$ in (3.7). In the case where det$\P^\#(\hat A)=1$,
their $N$-th powers are
$$
\P^\#(\w^{-1/2})^N=(b^\#/c^\#)^{N(N+1)/2}/H(a^\#/c^\#),       \eqno(3.12)
$$
where
$$
H(x)=(1-x)^N\prod_{j=1}^{N-1}(1-x\w^j)^j.                   \eqno(3.13)
$$
In the basis, where operator $\hat U$ is diagonal, one can show that
relation (3.9) is nothing else but the restricted star-triangle relation
introduced by Bazhanov and Baxter [8] (see formulae (1.19), (1.20)
in their paper).

Let us comment on what is the ``classical'' limit of (3.9). Clearly, it
should correspond to $N\to\infty$. Asymptotics (1.5) for the function
(3.5) take the form:
$$
w(a,b,c|m)=\sqrt{1-x\over1-xt_m}\exp({1\over\epsilon}
[L_2(xt_m)-L_2(x)])(1+{\cal O}(\epsilon)),                   \eqno(3.14a)
$$
where
$$
\epsilon=2\pi i/N,\quad  x=a/b,\quad |a|< |b|,          \eqno(3.14b)
$$
and
$$
t_m=\lim_{N\to\infty}\w^m,\quad m\to\infty,\quad m/N={\rm fixed}.\eqno(3.14c)
$$
To define
the ``symbol'' of a finite dimensional operator, in analogy with
Section 2, consider two basis sets $\{|n\rangle \}$ and $\{|m\rangle \}$,
$n,m\in Z_N$, which
diagonalize operators $\hat U$ and $\hat V$, respectively:
$$
\langle n|\hat U|n'\rangle =\w^{n-1/2}\delta_{n,n'},\quad
\langle n|\hat V|n'\rangle =\w^{-1/2}\delta_{n,n'+1},
\eqno(3.15a)
$$
$$
\langle m|\hat V|m'\rangle =\w^{m-1/2}\delta_{m,m'},\quad
\langle m|\hat U|m'\rangle =\w^{-1/2}\delta_{m,m'-1}.  \eqno(3.15b)
$$
The scalar products are given by
$$
\langle n|n'\rangle =\delta_{n,n'},\quad
\langle n|m\rangle =\w^{-nm},\quad\langle m|m'\rangle =N\delta_{m,m'},
   \eqno(3.16)
$$
while the decompositions of the unit operator are
$$
1=\sum_{n\in Z_N}|n\rangle \langle n|
={1\over N}\sum_{m\in Z_N}|m\rangle \langle m|.         \eqno(3.17)
$$
The ``U-V'' symbol of an operator, $\hat A$, is defined as a mixed matrix
element
$$
(\hat A)_{n,m}={\langle n|\hat A|m\rangle \over\langle n|m\rangle },
\eqno(3.18)
$$
the symbol of a product of $\hat A$ and $\hat B$ being
$$
(\hat A\hat B)_{n,m}={1\over N}\sum_{n',m'\in Z_N}\w^{-(n-n')(m'-m)}
(\hat A)_{n,m'}(\hat B)_{n',m}.           \eqno(3.19)
$$
Calculating the symbol of (3.9), we will have a direct analog for (2.16) with
a double sum instead of integrals. To apply the stationary phase method, we
have
to convert these summations into integrals.
For this purpose, first let us write the following equality:
$$
\sum_{m=0}^{N-1}{1\over z-\w^m}={N\over z(1-z^{-N})}   \eqno(3.20)
$$
for any complex $z$. Then, using Cauchy's formula and (3.20), we have
with exponential accuracy
$$
\sum_{m=0}^{N-1}f(\w^m)\sim{N\over2\pi i}\oint_{S^1}f(z){dz\over z},
\quad N\to\infty                                   \eqno(3.21)
$$
for any function $f(z)$. Repeating calculations from Section 2, using (3.21),
one can show that (3.9) in the limit $N\to\infty$ is reduced to $(1.3b)$
as well.

\beginsection{Summary}

In this paper we have shown that the quantum identity (2.4) for the
function (1.4) with operator arguments satisfying (2.1) in the
classical limit is reduced to
Rogers' dilogarithm identity (1.3). We used the ``q-p'' symbols
(2.10), (2.12) of the
operators $\hat U$, $\hat V$, and $-\hat U\hat V$
to convert (2.4) into the integral identity (2.16), as well as
asymptotics (1.5), and
the stationary phase method to evaluate the integrals.

In Section 3 we proved another, finite dimensional counterpart of the
quantized Rogers' identity (3.9).
The deformation parameter in this case is a root
of unity $\t^N=1$ for some integer $N\ge2$. The quantum dilogarithm is a
function on Fermat curve and is given explicitly by
(3.4), (3.5), (3.7), and (3.12). The classical limit of (3.9), corresponding to
$N\to\infty$, also coincides with (1.3).

It  is remarkable that the five term identity, which can be shown to be
equivalent to (3.9), but written in different terms, has been found already by
Bazhanov and Baxter [8].
They used it as a local integrability condition for three dimensional
statistical models.
An interesting question in connection with this is whether one can
supply a three dimensional interpretation for the quantum dilogarithm identity.
 In fact, the answer to this question is affirmative, and it will be the
subject of a subsequent publication.

\beginsection{References}

\noindent
\item{[1]}{H.W.J. Bl\"ote, J.L. Cardy
and M.P. Nightingale,  Phys. Rev. Lett. {\bf 56} (1986) 742}
\item{}{A.B. Zamolodchikov, Int. J. Mod. Phys. A{\bf4} (1989) 4235}
\item{}
{Al.B. Zamolodchikov, Nucl. Phys. B{\bf342} (1990) 695;
B{\bf358} (1991) 497}
\item{}
{V.V. Bazhanov and N.Yu. Reshetikhin, Prog. Theor. Phys. Suppl.{\bf 102}
(1990) 301}
\item{}{M.J. Martins, Phys. Rev. Lett. {\bf 65} (1990) 2091}
\item{}{T.R. Klassen and E. Melzer, Nucl. Phys. B{\bf338} (1990) 485;
B{\bf350} (1991) 635}
\item{}{V.A. Fateev and Al.B. Zamolodchikov, Phys. Lett. B{\bf271}
(1991) 91}
\item{[2]}{H.M. Babujan, Nucl. Phys. B{\bf215}[FS7] (1983) 317;}
\item{}{I. Affleck, Phys. Rev. Lett.{\bf 56} (1986) 746}
\item{}{H.J. de Vega and M. Karowski, Nucl. Phys. B{\bf285}[FS19] (1987)
619;}
\item{}{A.N. Kirillov and N.Yu. Reshetikhin, J. Phys. A{\bf 20} (1987)
1587}
\item{}{V.V. Bazhanov and Yu.N. Reshetikhin, Int. J. Mod. Phys. A{\bf 4}
(1989) 115}
\item{}{V.V. Bazhanov and Yu.N. Reshetikhin, J. Phys. A{\bf 23} (1990)
1477}
\item{}{A. Kl\"umper, M.T. Batchelor and P.A. Pearce, J. Phys. A{\bf
24}
(1991)
3111}
\item{}{A. Kl\"umper and P.A. Pearce, ``Conformal weights of RSOS
lattice
models
and their fusion hierarchies", preprint No. 23-1991 (1991)}
\item{}{A. Kuniba, ``Thermodynamics of the $U_q(X_r^{(1)})$ Bethe Ansatz
System
With $q$ A Root of Unity", ANU preprint (1991)}
\item{[3]}{R.J. Baxter, Physica {\bf 18D} (1986) 321}
\item{[4]}{L.J. Rogers, Proc. London Math. Soc. {\bf4} (1907) 169-189}
\item{[5]}{L. Faddeev, L. Takhtajan, in: Lecture notes in physics, Vol.
246 (Springer, Berlin, 1986) p.66}
\item{}{A.Yu. Volkov, Zap. Nauchn. Semin. LOMI 161 (1987) 24 [transl. in
J. Sov. Math.]}
\item{}{A.Yu. Volkov, Theor. Math. Phys. 74 (1988) 135}
\item{}{O. Babelon, Phys. Lett. B{\bf238} (1990) 234}
\item{}{A.Yu. Volkov, Phys. Lett. A{\bf167} (1992) 345}
\item{[6]}
{L. Faddeev and A.Yu. Volkov,``Abelian Current Algebra and the
Virasoro Algebra on the Lattice'', Preprint HU-TFT-93-29}
\item{[7]}
{Berezin, F.A., ``The Method of Second Quantization'' (in Russian),
Nauka, Moscow, 1965; English transl., Academic Press, New York, 1966.
\item{[8]}{V.V. Bazhanov, R.J. Baxter, J. Stat. Phys. {\bf71} (1993)
839}
\bye